\documentclass[reprint,prl,amsmath,amssymb,aps,showpacs,preprintnumbers]{revtex4-1}
\usepackage{graphicx} % Include figure files
\usepackage{datetime}
\usepackage{bm} % bold math

\begin{document}
\author{Tigran V. Shahbazyan}
%\email{shahbazyan@jsums.edu}
\affiliation{Department of Physics, Jackson State University, Jackson, MS 39217, USA}

\title{Exciton-plasmon energy exchange drives the transition to a strong coupling regime}

\begin{abstract} 
We present a model for exciton-plasmon coupling based on an energy exchange mechanism between quantum emitters (QE) and localized surface plasmons in metal-dielectric structures. Plasmonic correlations between QEs give rise to a collective state  exchanging its energy cooperatively with a resonant plasmon mode. By defining carefully the plasmon mode volume for a QE ensemble, we obtain a relation between QE-plasmon coupling and a cooperative energy transfer rate that is expressed in terms of local fields. For a single QE near a sharp metal tip, we find analytically the enhancement factor for   QE-plasmon coupling relative to  QE  coupling to a cavity mode. For QEs distributed in an extended region enclosing a plasmonic structure, we find that  the ensemble QE-plasmon coupling saturates to a universal value independent of system size and shape, consistent with the experiment. 
\end{abstract}
%{\bf Keywords:} surface plasmons, strong coupling, excitons, energy transfer, Rabi splitting
\maketitle

%%%%%%%%%%%%%%%%%%%%%%%%%%%%%%%%%%%%%%%%%%%%
\clearpage
%
%%%%%%%%%%%%%%%%%%%%%%%%%%%%%%%%%%%%%%%%%%
%\section{Introduction}
%\label{sec:intro}

Strong coupling of localised surface plasmons in metal-dielectric structures with excitons in dye molecules or semiconductors attracts intense interest due to emerging applications such as  ultrafast reversible switching \cite{ebbesen-prl11,bachelot-nl13,zheng-nl16}, quantum computing \cite{waks-nnano16,senellart-nnano17}, and light harvesting \cite{leggett-nl16}. An extremely strong field confinement on the length scale well below the diffraction limit provided by plasmonic structures gives rise to exciton-plasmon coupling that is much stronger than the exciton coupling to cavity modes in  semiconductor microcavites.  A strong coupling regime between two systems is established when the energy is exchanged between them faster than it dissipates,  which leads to an anticrossing gap   (Rabi splitting) in the dispersion of mixed states \cite{novotny-book}. While  relatively weak Rabi splittings  $\sim 1$ meV were reported for semiconductor quantum dots (QDs) coupled to a cavity mode \cite{forchel-nature04,khitrova-nphys06,imamoglu-nature06},  much greater splittings (up to 500 meV) were observed for surface plasmons  coupled  to excitons in J-aggregates \cite{bellessa-prl04,sugawara-prl06,wurtz-nl07,fofang-nl08,lienau-prl08,bellessa-prb09,schlather-nl13,lienau-acsnano14,shegai-prl15,shegai-nl17}, various dye molecules \cite{hakala-prl09,berrier-acsnano11,salomon-prl12,luca-apl14,noginov-oe16},   semiconductors QDs \cite{gomez-nl10,gomez-jpcb13}, or  two-dimensional atomic crystalls \cite{xu-nl17,alu-oe18,urbaszek-nc18,shegai-nl18}.

%%%%%%%%%%%%%%%%%%%%%%%%%%%%%%%%%%%%%%%%%%%%%%%%%%%%%%%%%%%

In  the classical picture of coupled oscillators  \cite{novotny-book}, the interaction between a quantum emitter (QE), modeled here by a two-level system  with frequency $\omega_{0}$, and a cavity (or plasmon) mode with frequency $\omega_{m}$, gives  rise to mixed states that show up in optical spectra through splitting of scattering or emission peaks into  two polaritonic bands  separated (for $\omega_{m}=\omega_{0}$) by  $\Delta=\sqrt{4g^{2}-(\gamma_{m}-\gamma_{0})^{2}/4}$, where  $\gamma_{0}$ and $\gamma_{m}$ are, respectively, the QEs and mode decay rates and $g$ is the coupling parameter \cite{khitrova-nphys06},
\begin{equation}
\label{coupling}
g=  \sqrt{ \frac{2\pi \mu^{2}\omega_{m}}{\hbar{\cal V}}}.
\end{equation}
Here $\mu$ is  the QE dipole matrix element  and ${\cal V}$ is the mode volume characterizing field confinement at the QE position.  Since typically $\gamma_{0}\ll \gamma_{m}$, the onset of transition to the strong coupling regime is $g\simeq \gamma_{m}/4$, which demands sufficiently strong field confinement (or small mode volume). For a QD placed inside  a microcavity, the cavity mode volume is at least ${\cal V}_{\rm cav}\sim(\lambda/2)^{3}$, where $\lambda$ is the wavelength, resulting in  $\Delta\sim 1$ meV  \cite{khitrova-nphys06}. However, if a QE is located at a  "hot spot" near a plasmonic structure characterized by extremely strong field confinement, e.g., near a sharp metal tip or in a gap between two metal structure, the Rabi splitting can reach much greater values even though the plasmon decay rate is much higher than that for cavity modes \cite{haran-nc16,baumberg-nature16,lienau-acsph18,pelton-nc18}.

If an ensemble of $N$ QEs is coupled to a cavity mode,  a collective state is formed that is comprised of  all QEs oscillating in sync with each other \cite{khitrova-rmp99}.  For such a  collective state, the coupling  scales as $g_{N}\propto \sqrt{N}$ with the ensemble size, while the rest of collective states are "dark"; i.e., they interact  only weakly with the  electromagnetic field. Importantly, such scaling implies that individual QEs are  coupled to the cavity mode with approximately equal strength, so that the mode, in fact, interacts with one "giant oscillator" whose amplitude is enhanced by the factor $N$ relative to individual QE.

\begin{figure}[tb]
%\centering
\begin{center}
\includegraphics[width=0.9\columnwidth]{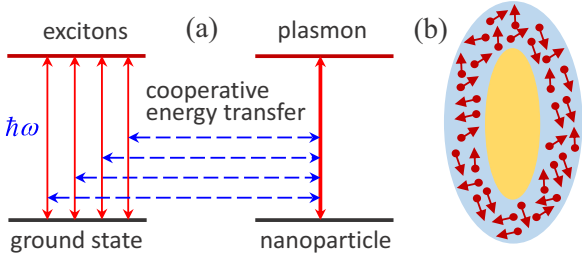}
\caption{\label{fig1} (a) Schematic  of exciton-plasmon coupling mediated by cooperative energy transfer between a collective state and resonant plasmon mode. (b) Schematic  of an open plasmonic system with QEs embedded in dielectric shell enclosing a metallic core.}
\end{center}
\vspace{-5mm}
\end{figure}

This picture is affected dramatically if an ensemble of $N$ QEs is placed near a plasmonic nanostructure, which is normally characterized by strongly varying local fields. For example, in a typical experimental setup, the QEs (excitons in J-aggregates or QDs) are embedded  in a dielectric shell enclosing metallic core supporting a localized surface plasmon (see Fig. \ref{fig1}). Strong exciton-plasmon coupling requires large plasmon local density of states (LDOS) facilitating an efficient energy transfer (ET) between a plasmon and a QE. In plasmonic structures,  local  fields can be very strong close to the metal surface, especially near sharp features such as narrow tips or surface irregularities, but fall  off rapidly away from the   surface. In this case, the coupling between individual QEs and resonant plasmon mode can vary in a wide range, so that the classical "giant oscillator" picture is no longer useful  and, instead, one has to resort to the underlying mechanism of energy exchange between the plasmon and QEs. Furthermore, since  for sufficiently remote QEs, the individual QE-plasmon ET rates are small, it is evident that the ensemble QE-plasmon coupling should \textit{saturate} as the region, in which the QEs are distributed, expands. Saturation of Rabi splitting was recently reported for  molecular excitons in J-aggregates embedded in dielectric shell enclosing Au nanoprism \cite{shegai-nl17}.

%%%%%%%%%%%%%%%%%%%%%%%%%%%%%%%%%%%%%%%%%%%%%%
%
%
%%%%%%%%%%%%%%%%%%%%%%%%%%%%%%%%%%%%%%%%%%%%%

In this Letter, we develop a model for exciton-pasmon coupling based upon microscopic picture of energy exchange  between the system components. We establish an explicit relation between the ensemble QE-plasmon coupling and the rate of cooperative energy transfer (CET) between a collective state and resonant plasmon mode (see Fig. \ref{fig1}), and estimate the energy-exchange frequency in the strong coupling regime. By defining carefully the plasmon mode volume in terms of the  plasmon LDOS, we show that, for a single QE at a hot spot near a sharp metal tip,  the QE-plasmon coupling scales as $g\propto V_{\rm met}^{-1/2}$, where $V_{\rm met}$ is the metal volume fraction that confines the plasmon field, and provide an analytical expression for the enhancement factor relative to the exciton coupling to a cavity mode. For an ensemble of QEs near a plasmonic structure, we show that the ensemble coupling parameter scales as $g_{N}\propto (N/\bar{\cal V})^{1/2}$, where $\bar{\cal V}$ is the \textit{average} plasmon mode volume in the QE region. If QEs are placed in a region with nearly constant, however large, plasmon LDOS, the coupling exhibits the usual cavity-like  scaling $g_{N}\propto \sqrt{N}$. However, in \textit{open} plasmonic systems, if QEs are uniformly distributed, with concentration $n$, in an extended region outside the plasmonic structure (see Fig. \ref{fig1}),   the  ensemble QE-plasmon coupling saturates to some value that, for system size below the diffraction limit, has a universal form
\begin{equation}
\label{coupling-saturated0}
g_{s}
=\sqrt{\frac{4\pi\mu^{2}n|\varepsilon'(\omega_{m})|}{3\hbar \varepsilon_{d}\partial \varepsilon'(\omega_{m})/\partial \omega_{m}}},
\end{equation}
where $\varepsilon(\omega)=\varepsilon'(\omega)+i\varepsilon''(\omega)$ is the metal dielectric function and $\varepsilon_{d}$ is the dielectric constant in the QE region. The saturated coupling $g_{s}$ is independent of system's size and shape, except indirectly via the plasmon frequency $\omega_{m}$. Surprisingly, $g_{s}$ is comparable to the coupling between QEs, uniformly distributed   inside a microcavity,  and a cavity mode, suggesting that  large Rabi splittings, observed in open plasmonic systems, are likely due to high QE concentrations in such systems.

\section{Plasmon LDOS, Mode Volume and Energy Transfer Rate}
\label{plasmon}

To establish a relation between cavity-like coupling (\ref{coupling})  and microscopic picture of energy exchange in plasmonic systems, we employ a classical approach for plasmons and describe excitons by  two-level  QEs with dipole moment $\bm{\mu}=\mu \bm{n}$, where $\bm{n}$ is the dipole orientation.  Note that although, in contrast to excitons, two-level QEs are Fermionic systems, this distinction plays no role in the QE-plasmon energy exchange processes considered here as long as the exciton-exciton (or QE-QE) interactions are relatively weak. We assume that QEs  are placed near a metal-dielectric  structure with characteristic size smaller than the radiation wavelength described by a complex  dielectric function $\varepsilon (\omega,\bm{r})=\varepsilon' (\omega,\bm{r})+i\varepsilon'' (\omega,\bm{r})$. The interaction of a QE, positioned at $\bm{r}$,  with electromagnetic environment  is characterized by the projected LDOS \cite{novotny-book}
\begin{equation}
\label{ldos}
\rho_{n}(\omega,\bm{r})=\frac{1}{2\pi^{2} \omega} \, \text{Im} [\bm{n} \bar{\bm{D}}(\omega;\bm{r},\bm{r})\bm{n}],
\end{equation}
where $\bar{\bm{D}}(\omega;\bm{r},\bm{r}')=4\pi k^{2} \bar{\bm{G}}(\omega;\bm{r},\bm{r}')$ is the rescaled electromagnetic dyadic Green function in the presence of plasmonic structure ($\bar{\bm{G}}$ is the standard Green dyadic for Maxwell equations and $k$ is the wave vector). In the near field limit, the Green function splits into free-space and plasmon contributions, $\bar{\bm{D}}=\bar{\bm{D}}_{0}+\bar{\bm{D}}_{m}$, where the plasmon Green function, for frequency $\omega$ near the plasmon mode frequency $\omega_{m}$, has the form \cite{shahbazyan-prl16,shahbazyan-prb18}:
\begin{equation}
\label{dyadic-mode}
\bar{\bm{D}}_{m}(\omega;\bm{r},\bm{r}') = \frac{\omega_{m}}{4 U_{m}}
\frac{\bm{E}(\bm{r}) \bm{E} (\bm{r}')}{\omega_{m}-\omega -i\gamma_{m}/2}.
\end{equation}
Here, $\bm{E}(\bm{r})$ is the plasmon field, which we chose to be real,   satisfying the Gauss's law, 
$\bm{\nabla}\!\cdot\! \left [\varepsilon' (\omega_{m},\bm{r}) \bm{E}(\bm{r})\right ]=0$, and the tensor product of fields is implied. In Eq.~(\ref{dyadic-mode}), $U_{m}$ is the plasmon mode energy defined in the standard way as \cite{landau},
\begin{align}
\label{energy-mode}
U_{m}
= \frac{1}{16\pi} 
\!\int \!  dV \, \frac{\partial [\omega_{m}\varepsilon'(\omega_{m},\bm{r})]}{\partial \omega_{m}} \bm{E}^{2}(\bm{r}),
\end{align}
while the factor $\omega_{m}/4U_{m}$ is the plasmon pole residue in the complex frequency plane \cite{shahbazyan-prb18}. Using Eqs.~(\ref{ldos}) and (\ref{dyadic-mode}), we obtain the projected \textit{plasmon} LDOS,
\begin{equation}
\label{ldos-pl}
\rho_{n}(\omega,\bm{r})
=\frac{Q}{4\pi^{2} \omega U_{m}}\, 
\frac{[\bm{n}\!\cdot\!\bm{E}(\bm{r})]^{2}}{1+4Q^{2}(\omega/\omega_{m}-1)^{2}},
\end{equation}
where $Q=\omega_{m}/\gamma_{m}$ is the plasmon quality factor. The full plasmon LDOS \cite{shahbazyan-prl16,shahbazyan-prb18}  $\rho (\omega,\bm{r})$ is obtained from Eq.~(\ref{ldos-pl}) with the replacement $[\bm{n}\!\cdot\!\bm{E}(\bm{r})]^{2}\rightarrow \bm{E}^{2}(\bm{r})$. The plasmon LDOS  describes the distribution of plasmon states in   unit volume and frequency interval. Accordingly, its frequency integral, $\rho (\bm{r})=\!\int\! d\omega \rho (\omega,\bm{r})$, defines the plasmon mode density, which, in the projected case, has the form \cite{shahbazyan-prl16,shahbazyan-prb18}
\begin{equation}
\label{density}
\rho_{n} (\bm{r})
%=\frac{[\bm{n}\!\cdot\!\bm{E}(\bm{r})]^{2}}{8\pi U }
= \frac{2[\bm{n}\!\cdot\!\bm{E}(\bm{r})]^{2}}{\int \! dV [\partial (\omega_{m}\varepsilon')/\partial \omega_{m}]\bm{E}^{2}} 
= \frac{1}{{\cal V}_{n}},
\end{equation}
where ${\cal V}_{n}$  is the projected \textit{plasmon mode volume} that characterizes the plasmon field confinement at a point $\bm{r}$ in the direction $\bm{n}$. Note that ${\cal V}_{n}$ is a \textit{local} quantity that can vary significantly near a plasmonic structure.
%, and which, according to Eq.~(\ref{coupling}), determines  the QE-plasmon coupling. 
We stress that, for characteristic system size smaller than the radiation wavelength, the plasmon mode volume is a  \textit{real} function of $\bm{r}$ defined in terms of real plasmon fields at that position \cite{shahbazyan-prl16,shahbazyan-prb18}. For larger systems, where plasmons are strongly hybridized with the radiation continuum,  field confinement is not a well-defined notion and, in this case, QE-plasmon interactions can be described using the expansion over quasi-normal modes  characterized by complex mode volumes \cite{lalanne-prl13,hughes-njp14,hughes-acsphot14,muljarov-prb16,lalanne-lpr18}. In our approach, the above connection between the  plasmon mode volume and LDOS allows relating the  QE-plasmon coupling to the QE-plasmon ET rate.  

In order to relate   ${\cal V}_{n}$  to the QE-plasmon ET rate $\gamma_{et}$, we note that the latter can be expressed via the plasmon Green function as \cite{shahbazyan-prl16,shahbazyan-prb18}  $\gamma_{et}(\omega)=(2\mu^{2}/\hbar) \, \text{Im} [\bm{n}\bar{\bm{D}}_{m}(\omega;\bm{r},\bm{r})\bm{n}]$. Using Eq.~(\ref{dyadic-mode}), we obtain the frequency-dependent ET rate as a Lorentzian centered at the plasmon frequency,
\begin{equation}
\label{rate-mode}
\gamma_{et}(\omega)
=\frac{\mu^{2}Q}{\hbar U_{m}}\,\frac{[\bm{n}\!\cdot\!\bm{E}(\bm{r})]^{2}}{1+4Q^{2}(\omega/\omega_{m}-1)^{2}}.
%=\frac{8\pi\mu^{2}Q}{\hbar {\cal V}_{n}}.
%=\frac{16\pi\mu^{2}Q[\bm{n}\!\cdot\!\bm{E}(\bm{r})]^{2}}{\hbar\int \! dV [\partial (\omega_{m}\varepsilon')/\partial \omega_{m}]\bm{E}^{2}}.
\end{equation}
At resonance frequency $\omega=\omega_{m}$, the QE-plasmon ET rate $\gamma_{et}\equiv\gamma_{et}(\omega_{m})$ takes the form
\begin{equation}
\label{rate-mode-res}
\gamma_{et}
%=\frac{\mu^{2}Q}{\hbar U}\, [\bm{n}\!\cdot\!\bm{E}(\bm{r})]^{2}
=\frac{16\pi\mu^{2}Q[\bm{n}\!\cdot\!\bm{E}(\bm{r})]^{2}}{\hbar\int \! dV [\partial (\omega_{m}\varepsilon')/\partial \omega_{m}]\bm{E}^{2}}
=\frac{8\pi\mu^{2}Q}{\hbar {\cal V}_{n}},
\end{equation}
where ${\cal V}_{n}$ is given by Eq.~(\ref{density}). Normalizing $\gamma_{et}$ by  the free-space radiative decay rate $\gamma_{0}^{r}=4\mu^{2}k^{3}/3\hbar$, we recover the Purcell factor \cite{purcell-pr46}  for a QE coupled to a plasmonic resonator: $F_{p}=\gamma_{et}/\gamma^{r}_{0}=6\pi Q/{\cal V}_{n}k^{3}$.

%%%%%%%%%%%%%%%%%%%%%%%%%%%%%%%%%%%%%%%%%%%%%%
%
\begin{figure}[tb]
%\centering
\begin{center}
\includegraphics[width=0.9\columnwidth]{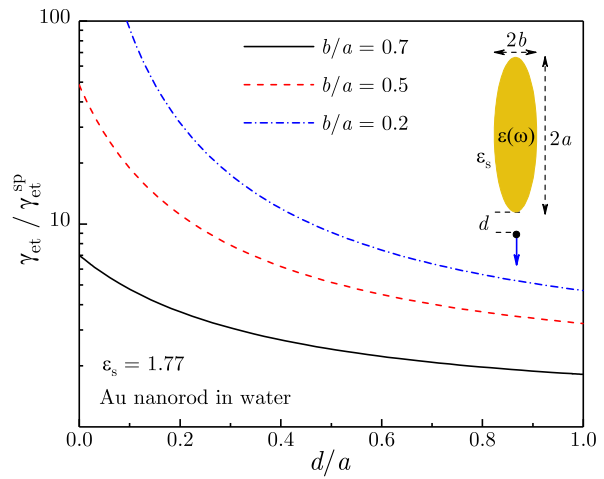}
\caption{\label{fig2} Normalized QE-plasmon ET rate  plotted against the QE distance from  Au nanorod tip for several nanorod aspect ratios. Inset. Schematic  of a QE placed at a distance $d$ from a tip of Au nanorod in water.}
\end{center}
\vspace{-5mm}
\end{figure}
%
%%%%%%%%%%%%%%%%%%%%%%%%%%%%%%%%%%%%%%%%%%%%%

To illustrate  the ET rate's sensitivity to the QE position and system geometry, in Fig.~\ref{fig2} we show  $\gamma_{et}$ for a QE placed at a distance $d$ from the tip of an Au nanorod in water.  Nanorod was modeled by a prolate speroid with semimajor and semiminor axes $a$ and $b$, respectively, and the experimental Au dielectric function was used in all calculations.  To highlight the dependence of $\gamma_{et}$ on system geometry, we normalize it by  its value $\gamma_{et}^{sp}$ for a sphere with radius $a$, and plot the result   against the normalized distance $d/a$ for several aspect ratios $a/b$. With increasing aspect ratio, i.e., reducing the nanorod volume relative to the sphere, the normalized ET rates increase dramatically near the nanorod tip, which translates to analogous behavior of the QE-plasmon coupling, as we discuss in detail later in the paper.

Let us now turn to an  \textit{ensemble} of $N$ QEs with dipole moments $\bm{p}_{i}=\mu \bm{n}_{i}$ situated at  positions $\bm{r}_{i}$ near a  plasmonic  structure. Each QE interacts, via the coupling  $\bm{p}_{i} \cdot \bm{\mathcal{E}}(\bm{r}_{i})$, with the common electric field $\bm{\mathcal{E}}(\omega,\bm{r} ) = \sum_{j} \bar{\bm{D}}(\omega;\bm{r},\bm{r}_{j}) \bm{p}_{j}$ generated by all QEs, where $\bar{\bm{D}}(\omega;\bm{r},\bm{r}_{j})$ is the near-field electromagnetic Green's function in the presence of plasmonic structure. Due to electromagnetic correlations between QEs, collective states are formed that are described by the eigenstates of the Green's function matrix at QEs' positions projected onto QEs' dipole moments \cite{haroche-pr82}: $D_{ij}=\bm{p}_{i} \bar{\bm{D}}(\omega;\bm{r}_{i},\bm{r}_{j})  \bm{p}_{j}$. Note that this coupling matrix  includes direct dipole-dipole interactions between the QEs which are causing random shifts of the QEs' energies. However, for random dipole orientations, the direct dipole coupling  vanishes \textit{on average} \cite{friedberg-pr73,shahbazyan-prb00}, while its  fluctuations contribute, among other factors, to inhomogeneous broadenning of the QE energies (we  return to this point later). Furthermore, for system size smaller than the radiation wavelength, the radiative (superradiant) coupling between the QEs is weak as compared to the resonant plasmon-assisted coupling, and is not considered here. Therefore, near the plasmon resonance, the  coupling matrix is dominated by the plasmon Green function Eq.~(\ref{dyadic-mode}), and we obtain \cite{shahbazyan-prb19}
\begin{equation}
\label{matrix-pl}
D_{ij}^{m}(\omega)=\frac{\omega_{m}}{4 U_{m}}\frac{\bm{p}_{i}\!\cdot\!\bm{E}(\bm{r}_{i})\, \bm{p}_{j}\!\cdot\!\bm{E}(\bm{r}_{j})}{\omega_{m}-\omega-i\gamma_{m}/2}.
\end{equation}
The diagonal elements of plasmon coupling matrix Eq.~(\ref{matrix-pl}) are complex, $D_{ii}=\hbar\delta\omega^{(i)}+i\hbar \gamma_{\rm et}^{(i)}/2$, and their real and imaginary parts describe, respectively, the frequency shifts $\delta\omega^{(i)}$ and decay rates $\gamma_{\rm et}^{(i)}$ of individual QEs due to interaction with a plasmon mode; the former contribute to inhomogeneous broadening of the QE energies, while the latter are the QE-plasmon ET rates, given by  Eq.~(\ref{rate-mode}). Note that at resonance, we have $\delta\omega^{(i)}=0$, while $\gamma_{\rm et}^{(i)}$ are given by  Eq.~(\ref{rate-mode-res}). 

The off-diagonal elements of plasmon coupling matrix Eq.~(\ref{matrix-pl}) describe plasmonic correlations between QEs  which give rise to  collective states. These states are represented by the eigenstates $\psi$ of the coupling matrix (\ref{matrix-pl}) satisfying $\hat{D}\psi=\lambda\psi$, where the complex eigenvalues $\lambda$ characterize the energy shifts and decay rates of the collective states.  It is now easy to see that the vector $\psi_{N}=\{\bm{p}_{1}\!\cdot\!\bm{E}(\bm{r}_{1}),\dots,\bm{p}_{N}\!\cdot\!\bm{E}(\bm{r}_{N})\}$ is an eigenstate of the matrix Eq.~(\ref{matrix-pl}), and its eigenvalue has the form $\lambda_{N}=\hbar\delta\omega_{N}+i\hbar \gamma_{\rm et}^{N}/2$, where $\delta\omega_{N}=\sum_{i}\delta\omega^{(i)}$ is frequency shift of the collective state and $\gamma_{\rm et}^{N}
=\sum_{i}\gamma_{\rm et}^{(i)}$ is its decay rate. At resonance $\omega=\omega_{m}$, we again have $\delta\omega_{N}=0$, while the decay rate of the collective state $\psi_{N}$ takes the form [compare to Eq.~(\ref{rate-mode-res})]
\begin{equation}
\label{rate-cet}
\gamma_{et}^{N}=\!\sum_{i}\frac{16\pi\mu^{2}Q[\bm{n}_{i}\!\cdot\!\bm{E}(\bm{r}_{i})]^{2}}{\hbar\int \! dV [\partial (\omega_{m}\varepsilon')/\partial \omega_{m}]\bm{E}^{2}}
=\!\sum_{i}\frac{8\pi\mu^{2}Q}{\hbar {\cal V}_{n}^{(i)}},
\end{equation}
where ${\cal V}_{n}^{(i)}$ is the projected plasmon mode volume (\ref{density}) at the $i$th QE position. Thus, the collective state described by $\psi_{N}$ exchanges its energy with the plasmon mode \textit{cooperatively}, i.e., at a rate equal  to the \textit{sum} of individual QE-plasmon ET rates \cite{shahbazyan-prl16,shahbazyan-prb19}.  As we show below, the CET rate Eq.~(\ref{rate-cet}) determines onset of the transition to  strong coupling regime for QE ensembles coupled to plasmonic resonators.

Since the imaginary part of the eigenvalue $\lambda_{N}$ \textit{saturates}  the  ET rates from  QEs to  resonant plasmon mode, the rest of  collective  states are not coupled to that mode but, in principle, can still couple to  off-resonant modes and radiation field. Note, however, that for large ensembles,   coupling of collective states to  off-resonant modes is relatively weak \cite{vidal-prl14,petrosyan-prb17}, while large plasmonic Purcell factors ensure that coupling to the radiation field is relatively weak as well, implying that the exciton-plasmon energy exchange at the CET rate (\ref{rate-cet}) is the dominant energy flow channel in the system. We stress that, in plasmonic systems,  the collective states are formed due to QEs' correlations via the \textit{local} fields  that can vary strongly near a plasmonic structure.  Therefore, these states are \textit{distinct} from the superradiant and subradiant states, which emerge due to QEs' coupling to the common radiation field that is nearly uniform on the system scale.

\section{Exciton-Plasmon Coupling and Energy Exchange}

Consider first a single QE situated near a plasmonic resonator with its frequency $\omega_{0}$ close to the plasmon frequency $\omega_{m}$. We  consider systems with characteristic size smaller than the radiation wavelength and, therefore,  consider only the near-field coupling between a QE and resonant plasmon mode; note, however, that for larger systems,  radiative (superradiant) coupling between a QE and a plasmon  can significatly affect the optical spectra in the strong coupling regime \cite{lienau-acsnano14}. Using Eq.~(\ref{rate-mode}), we can now relate the QE-plasmon coupling (\ref{coupling}) to the QE-plasmon ET rate as
\begin{equation}
\label{coupling-et}
g^{2}
%=\frac{2\pi \mu^{2}\omega_{m}}{\hbar{\cal V}}
=\frac{1}{4}\gamma_{et}\gamma_{m}
=\frac{4\pi\mu^{2}\omega_{m}[\bm{n}\!\cdot\!\bm{E}(\bm{r})]^{2}}{\hbar\int \! dV [\partial (\omega_{m}\varepsilon')/\partial \omega_{m}]\bm{E}^{2}}.
\end{equation}
From Eq.~(\ref{coupling-et}), a relation between the QE-plasmon coupling and the Purcell factor follows: $g^{2}=F_{p}\gamma_{0}^{r}\gamma_{m}/4$.

Within classical model of coupled oscillators, the spectrum splits into upper and lower polaritonic bands with complex frequencies \cite{novotny-book} $\omega'_{\pm}=\frac{1}{2}(\omega'_{0}+\omega'_{m}\pm \sqrt{(\omega'_{0}-\omega'_{m})^{2}+4g^{2}})$, where  $\omega'_{0}=\omega_{0}-i\gamma_{0}/2$ and $\omega'_{m}=\omega_{m}-i\gamma_{m}/2$. To simplify the analysis, we assume for now that the QE and the plasmon are in resonance,  $\omega_{0}=\omega_{m}$. Typically,  in plasmonic systems, the intrinsic QE decay rate $\gamma_{0}$ is much smaller than either $\gamma_{m}$ or $\gamma_{et}$, and so we disregard it here as well; however small, $\gamma_{0}$ does play a crucial role in the interference effects, such as Fano resonances \cite{pelton-nc18}, but such phenomena are beyond our work's scope. In the weak coupling regime, both polaritonic bands are centered at the same frequency $\omega_{m}$ but are characterized by different spectral widths, 
\begin{equation}
\label{spectrum-weak}
\omega'_{\pm}=\omega_{m}-\frac{i}{4}\left (\gamma_{m} \pm \sqrt{\gamma_{m}^{2}-4\gamma_{m}\gamma_{et}}\right ),
\end{equation}
where we used the relation (\ref{coupling-et}). For $\gamma_{et}\ll \gamma_{m}$, the expansion of Eq.~(\ref{spectrum-weak}) over the small parameter $\gamma_{et}/\gamma_{m}$ yields $\gamma_{+}= \gamma_{m} - \gamma_{et}$ and $\gamma_{-}=  \gamma_{et}$, which represent, respectively,  the decay rates of weakly-mixed plasmon and QE states. 

In the strong coupling regime, both polaritonic bands are characterized by constant spectral width $\gamma_{m}/2$, while their central frequencies are separated by the Rabi splitting: 
\begin{equation}
\label{spectrum-strong}
\omega'_{\pm}=\omega_{m}\pm\frac{1}{2}\sqrt{\gamma_{m}\gamma_{et}-\gamma_{m}^{2}/4} -i\frac{\gamma_{m}}{4}.
\end{equation}
Thus, in terms of QE-plasmon ET rate,  the transition  to strong coupling regime take place at 
\begin{equation}
\label{transition-single}
\gamma_{et}\simeq \gamma_{m}/4.
\end{equation}
Note that the polaritonic bands become spectrally distinct when Rabi splitting exceeds their linewidth $\gamma_{m}/2$, which corresponds to the condition $\gamma_{et}\gtrsim \gamma_{m}/2$.

The onset condition (\ref{transition-single}) implies that, for plasmonic resonators with high quality factor $Q=\omega_{m}/\gamma_{m}\gg 1$, the QE-plasmon energy exchange takes place on a much \textit{longer} time scale than the optical period. Qualitatively, the energy exchange dynamics can be described by the damped oscillator equation, $\ddot{x}+\gamma_{m}\dot{x}+\omega_{et}^{2}x=0$, where   $x$ characterizes the QE-plasmon energy balance while the frequency   $\omega_{et}=2\gamma_{et}$ is  the two-way QE-plasmon ET rate. After a pulsed excitation,  the energy balance between the QE and plasmon undergoes damped oscillations with energy-exchange frequency
\begin{equation}
\label{frequency-exchange}
\omega_{exc}=\sqrt{\omega_{et}^{2}-\gamma_{m}^{2}/4},
\end{equation}
and damping rate $\gamma_{m}/2$, which is similar to that for optical field \textit{intensity} in the strong coupling regime [see Eq.~(\ref{spectrum-strong})]. In this model, the transition between strong and weak coupling regimes is analogous to the transition between underdamped and overdamped  oscillator regimes taking place at $\omega_{et} =\gamma_{m}/2$, which coincides with Eq.~(\ref{transition-single}). 

Let us now turn to an ensemble of $N$ QEs near a plasmonic structure. As discussed in the previous section,   the plasmonic  correlations between QEs give rise to a collective state that exchanges its energy with the resonant plasmon mode at CET rate $\gamma_{et}^{N}$, given by (\ref{rate-cet}). Accordingly, the ensemble QE-plasmon  coupling $g_{N}$ is  related to the CET rate $\gamma_{et}^{N}$ as [compare to Eq.~(\ref{coupling-et})]
\begin{equation}
\label{coupling-cet}
g_{N}^{2}
=\frac{1}{4}\gamma_{et}^{N}\gamma_{m}
=\frac{2\pi\mu^{2}\omega_{m}N}{\hbar \bar{\cal V}},
\end{equation}
where $ \bar{\cal V} $ is the \textit{average} plasmon mode volume for QE ensemble, defined as 
\begin{equation}
\label{mode-volume-av}
\frac{1}{\bar{\cal V}}
=\frac{1}{N}\!\sum_{i} \frac{1}{{\cal V}_{n}^{(i)}} 
 =\frac{1}{N}\!\sum_{i}\! \frac{2[\bm{n}_{i}\!\cdot\!\bm{E}(\bm{r}_{i})]^{2}}{\int \! dV [\partial (\omega_{m}\varepsilon')/\partial \omega_{m}]\bm{E}^{2}}.
\end{equation}
The ensemble coupling (\ref{coupling-cet}) is related as $g_{N}^{2}=F_{p}^{N}\gamma_{0}^{r}\gamma_{m}/4$ to  cooperative Purcell factor $F_{p}^{N}=\gamma_{et}^{N}/\gamma_{0}^{r}=\sum_{i}F_{p}^{(i)}$ describing the enhancement of  CET-based cooperative emission rate \cite{shahbazyan-prb19}. The transition onset  to strong coupling regime is obtained by replacing the  ET rate  in   Eq.~(\ref{transition-single})  with the CET rate:  $\gamma_{et}^{N}\simeq \gamma_{m}/4$.  In the strong coupling regime, the energy balance between the collective state and resonant plasmon mode oscillates with energy-exchange frequency $\omega_{exc}^{N}=\sqrt{(2\gamma_{et}^{N})^{2}-\gamma_{m}^{2}/4}$,  which vanishes at the transition point [compare to Eq.~(\ref{frequency-exchange})].

The ensemble QE-plasmon coupling $g_{N}$ is related to  individual  QE couplings $g_{i}$ as $g_{N}^{2}=\sum_{i}g_{i}^{2}$  and, therefore, depends sensitively on the plasmon LDOS variations over the region QEs are distributed in. For example, if the plasmon LDOS (and, accordingly, the mode volume) is nearly constant in the QE region, e.g., within dielectric core enclosed by a metallic shell, the individual couplings $g_{i}$ are approximately equal and, in this case, the ensemble QE-plasmon coupling exhibits a cavity-like scaling, $g_{N} \propto \sqrt{N}$. However, in open plasmonic systems with  QEs distributed outside the plasmonic structure (see Fig.~\ref{fig1}), the ensemble QE-plasmon coupling saturates to a universal value, as we show later in this paper.

In the above analysis,  we assumed the same excitation frequencies for all QEs in the ensemble. In the experiment, however, the QE frequencies are distributed within some interval due to, e.g., direct dipole-dipole interactions between the QEs or, in the case of semiconductor QDs, their size variations. Here we note that the ET rate between a donor and an acceptor  is determined by the spectral overlap of their respective emission and absorption bands \cite{novotny-book}. Therefore, as long as the ensemble inhomogeneous broadening stays within the broad plasmon resonance band, the QE-plasmon energy exchange mechanism remains largely unaffected.

\section{Exciton-Plasmon Coupling  Near Sharp Metal Tip}

The largest values of exciton-plasmon coupling are achieved for QEs placed in a region with large plasmon LDOS that provides an efficient ET between QEs and a plasmon mode. Here, we consider a single QE situated near a sharp tip of a small plasmonic structure, such as a metal nanorod, where the field confinement can be extremely strong (hot spot). To estimate the QE-plasmon coupling, we note that the Gauss's law implies $\int \! dV [\partial (\omega_{m}\varepsilon')/\partial \omega_{m}]\bm{E}^{2}=\omega_{m}\partial \varepsilon'/\partial \omega_{m}\int \! dV_{\rm met}\bm{E}^{2}$, and so the  coupling  (\ref{coupling-et}) has the form 
\begin{equation}
\label{coupling-et2}
g^{2}
%=\frac{1}{4}\gamma_{et}\gamma_{m}
=\frac{4\pi\mu^{2}}{\hbar\partial \varepsilon'(\omega_{m})/\partial \omega_{m}}\frac{[\bm{n}\!\cdot\!\bm{E}(\bm{r})]^{2}}{\int \! dV_{\rm met} \bm{E}^{2}},
\end{equation}
where we assumed that only in the metallic region  is the dielectric function dispersive. The integral over  metallic region in Eq.~(\ref{coupling-et2}) depends on the characteristic size of that region $l_{m}$ relative to the skin penetration length $l_{s}$ (about 20 nm for Au in the plasmon frequency domain). For small structures with $l_{m}<l_{s}$, the QE-plasmon coupling scales as $g\propto V_{\rm met}^{-1/2}$, while in the opposite case,  $l_{m}>l_{s}$, the metal volume $V_{\rm met}$ should be replaced by the effective volume $V_{\rm met}^{\rm eff}\sim l_{s}^{3}$ that largely confines the plasmon field. In either case, there is a significant enhancement of $g$ relative to the QE coupling to a cavity mode,  $g_{\rm cav}\propto {\cal V}_{\rm cav}^{-1/2}\sim \lambda^{-3/2}$, and so in the analysis below we use the notation $V_{\rm met}$ for both cases. Importantly, in addition to this  geometric volume effect, there is  also strong field enhancement due to the "lightning rod" effect  near sharp metal tips or surface irregularities.

\begin{figure}[tb]
%\centering
\begin{center}
\includegraphics[width=0.9\columnwidth]{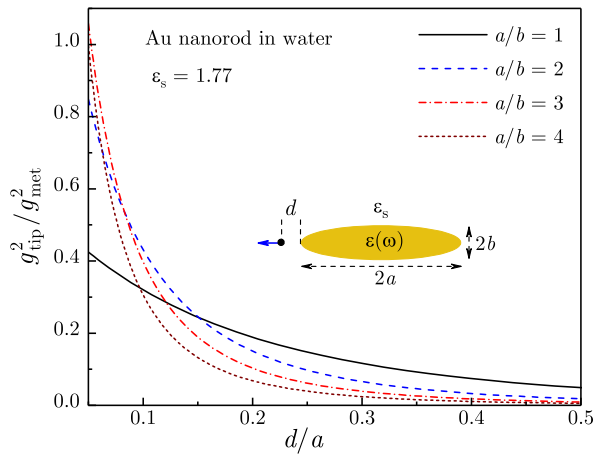}
\caption{\label{fig3} Normalized QE-plasmon coupling is plotted against the QE distance from Au nanorod tip for several nanorod aspect ratios.  Inset. Schematic of a QE placed at a distance $d$ from a tip of Au nanorod in water.
 }
\end{center}
\vspace{-4mm}
\end{figure}

To elucidate the relative importance of these two enhancement sources, below we estimate the coupling $g$ of a QE to a plasmon mode oscillating along the tip of a small metal structure (see Fig. \ref{fig3}). For small systems, i.e.,  $l_{m}<l_{s}$, the plasmon field  does  not vary significantly   inside the metallic structure while falling off rapidly  outside of it, so the largest field enhancement takes place near the tip. For a QE  polarized along  the tip  (see Fig.~\ref{fig3}), the coupling (\ref{coupling-et2}) along the system symmetry axis ($z$-axis) can be estimated as
\begin{equation}
\label{coupling-tip0}
g_{\rm tip}^{2}
=\frac{4\pi\mu^{2}}{\hbar V_{\rm met}\partial \varepsilon'(\omega_{m})/\partial \omega_{m}} 
\frac{ E^{2}(z) }{ E_{in}^{2}},
\end{equation}
where  the subscripts $in$ and $out$ refer, respectively, to the field values at the interface on  metal and dielectric side.  Using the boundary condition   $\varepsilon_{s}E_{out}=\varepsilon'(\omega_{m})E_{in}$, we obtain 
\begin{equation}
\label{coupling-tip}
g_{\rm tip}^{2}
=\frac{4\pi\mu^{2}}{\hbar V_{\rm met}}\frac{|\varepsilon'(\omega_{m})|^{2} \,\tilde{E}^{2}(z)}{\varepsilon_{s}^{2}\partial \varepsilon'(\omega_{m})/\partial \omega_{m}},
\end{equation}
where $\tilde{E}^{2}(z)$ is the plasmon field outside the structure normalized by its value at the tip. For good Drude metals, we have $\omega_{m}\partial \varepsilon'(\omega_{m})/\partial \omega_{m}\approx 2|\varepsilon'(\omega_{m})|$, and the coupling (\ref{coupling-tip}) takes the form
\begin{equation}
\label{coupling-tip1}
g_{\rm tip}^{2}
\approx \frac{2\pi\mu^{2}\omega_{m}|\varepsilon'(\omega_{m})|}{\hbar V_{\rm met}\varepsilon_{s}^{2}}\,\tilde{E}^{2}(z).
\end{equation}
The maximal value of $g_{\rm tip}$ is achieved at a close proximity to the tip, where $\tilde{E}\sim 1$. Comparing this value to the coupling $g_{\rm cav}$ for a QE inside a cavity, given by Eq.~(\ref{coupling}) with ${\cal V}\sim \lambda^{3}$, we obtain  the enhancement factor,
\begin{equation}
\label{enhancement-tip}
\frac{g_{\rm tip} }{g_{\rm cav} }\sim \sqrt{|\varepsilon'(\omega_{m})|\lambda^{3}/V_{\rm met}},
\end{equation}
indicating that the plasmonic enhancement is due to both the geometric volume effect and  the plasmon field enhancement characterized by   $|\varepsilon'(\omega_{m})|\gg 1$. Note that for metal nanostructures with characteristic size larger than the skin penetration length $l_{s}$, the metal volume in Eq.~(\ref{enhancement-tip}) should be replaced by $l_{s}^{3}$, as discussed above.

%%%%%%%%%%%%%%%%%%%%%%%%%%%%%%%%%%%%%%%%%%%%%%
%
%
%%%%%%%%%%%%%%%%%%%%%%

In Fig.~\ref{fig3}, we plot  distance dependence of the QE-plasmon coupling for a QE near the tip of an Au nanorod in water. Note that for a prolate spheroid used here to model the nanorod, the expression (\ref{coupling-tip}) is exact. The curves show calculated $g_{\rm tip}^{2}$, normalized by $g_{\rm met}^{2}=2\pi\mu^{2}\omega_{m}/\hbar V_{\rm met}$, plotted against normalized distance $d/a$ for different values of aspect ratio  $a/b$ at fixed nanorod length $a$, i.e., for different metal volumes $V_{\rm met}=ab^{2}$. It is clearly seen that for $a/b>1$, i.e., when the system possesses  hot spots near the nanorod tips, the QE-plasmon coupling scales as $V_{\rm met}^{-1/2}$  while falling off rapidly with increasing $d/a$. Note that,  while the coupling is largest at the tip ($\tilde{E}=1$),  it is expected to saturate  below  distances $d\sim v_{F}/\omega$ as the nonlocal effects become important \cite{smith-science12,mortensen-nc14,mortensen-acsphot18}. In  noble metals,  this length scale is $< 1$ nm in the plasmonic frequency range and, therefore, the distance dependence in Fig.~\ref{fig3} is cut off below $d/a=0.05$.

\section{Saturation of Exciton-Plasmon Coupling in Open Systems}

In open plasmonic systems, the QEs are distributed within some region \textit{outside} the metal structure, e.g., within  dielectric shell  enclosing a metallic core  [see Fig.~(\ref{fig1})]. In such systems, the plasmon field $\bm{E}(\bm{r})$ can vary substantially within the QE region and, in particular, falls off rapidly away from the metal surface. If QEs are uniformly distributed, with concentration $n$, within some volume $V_{0}$, the sum in  Eq.~(\ref{mode-volume-av}) can be replaced by  integral over $V_{0}$, and so the   corresponding QE-plasmon coupling $g_{0}$ takes the form
\begin{equation}
\label{coupling-uniform}
g_{0}^{2}
%=\frac{2\pi\mu^{2}\omega_{m}N}{\hbar {\cal V}_{0}}
=\frac{4\pi\mu^{2}n}{3\hbar \partial \varepsilon'(\omega_{m})/\partial \omega_{m}}
\frac{\int \! dV_{0}\bm{E}^{2}}{\int \! dV_{\rm met}\bm{E}^{2}},
\end{equation}
where the factor $1/3$ comes from orientational averaging. In contrast to the individual QE-plasmon coupling (\ref{coupling-et2}), the ensemble coupling (\ref{coupling-uniform}) is determined by the ratio of \textit{integrated} field intensities over the QE and metallic regions. 
If the QE region is sufficiently extended, the remote QEs do not interact with the  plasmon mode and, therefore, the integral $\int\! dV_{0}\bm{E}^{2}$ in Eq.~(\ref{coupling-uniform}) is  independent of $V_{0}$. In this case, the QEs saturate the entire plasmon mode volume, i.e., the plasmon LDOS  is  small beyond the QE region, and so  the  ensemble QE-plasmon coupling (\ref{coupling-uniform}) saturates to some value $g_{s}$. %Note that, since the coupling (\ref{coupling-uniform}) is independent of losses, its saturation is due to the reduction of QE-plasmon ET rates away from the plasmonic structure, rather than any dephasing processes.

Let us now show that the actual value of saturated coupling $g_{s}$ does \textit{not} depend directly on the size or shape of the metal structure as well and, remarkably, can be obtained explicitly for any plasmonic system with characteristic size below the diffraction limit. Indeed, due to the Gauss's law, both volume integrals in Eq.~(\ref{coupling-uniform}) reduce to surface integrals over the system interfaces, including the common metal-dielectric interface. After matching the normal field components at this interface and disregarding, in the saturated case, the outer boundary of the QE region, the ratio of integrated intensities in Eq.~(\ref{coupling-uniform}) is found as $\int\! dV_{0}\bm{E}^{2}/\int\! dV_{\rm met}\bm{E}^{2}=-\varepsilon'(\omega_{m})/\varepsilon_{d}$, and we arrive at a universal  saturated QE-plasmon coupling,
\begin{equation}
\label{coupling-saturated}
g_{s}^{2} 
=\frac{4\pi\mu^{2}n|\varepsilon'(\omega_{m})|}{3\hbar \varepsilon_{d}\partial \varepsilon'(\omega_{m})/\partial \omega_{m}}.
\end{equation}
Using   the relation $\omega_{m}\partial \varepsilon'(\omega_{m})/\partial \omega_{m}\approx 2|\varepsilon'(\omega_{m})|$ for good Drude metals,  the coupling (\ref{coupling-saturated})  simplifies to
\begin{equation}
\label{coupling-saturated2}
g_{s}  
\approx \sqrt{\frac{2\pi\mu^{2}n \omega_{m} }{3\hbar \varepsilon_{d} }},
\end{equation}
and  the transition onset to strong coupling regime for large ensembles, $g_{s}\simeq \gamma_{m}/4$,  can be presented as
\begin{equation}
\label{onset-saturated}
4Q\sqrt{\frac{2\pi\mu^{2}n  }{3\hbar \omega_{m} \varepsilon_{d}}}\simeq 1.
\end{equation}
Note that saturation of the QE-plasmon coupling in the strong coupling regime was recently reported for photolumenescence of molecular excitons in J-aggregates embedded in dielectric shell enclosing an Au nanoprism \cite{shegai-nl17}.

%%%%%%%%%%%%%%%%%%%%%%%%%%%%%%%%%%%%%%%%%%%%%%
%
\begin{figure}[bt]
%\centering
\begin{center}
\includegraphics[width=0.9\columnwidth]{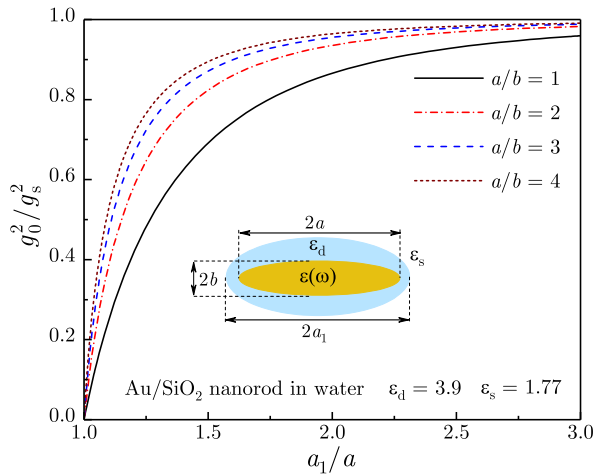}
\caption{\label{fig4} Calculated QE-plasmon coupling (\ref{coupling-uniform}) for QEs uniformly distributed within SiO$_2$ shell enclosing an Au nanorod normalized by saturated coupling  (\ref{coupling-saturated})
is plotted against  increasing shell thickness for several aspect ratios. Inset. Schematic of a core-shell Au/SiO$_2$ nanorod in water.
 }
\end{center}
\vspace{-4mm}
\end{figure}
%
%%%%%%%%%%%%%%%%%%%%%%

In Fig.~\ref{fig4}, we show the calculated QE-plasmon coupling $g_{0}$, given by Eq.~(\ref{coupling-uniform}), for QEs  distributed uniformly within a SiO$_2$ shell enclosing an Au nanorod. This core-shell system is modeled by two confocal spheroids with semimajor axes $a$ and $a_{1}$ corresponding to the Au/SiO$_2$ and Si$_2$O/H$_2$O interfaces, respectively. With  expanding QE region, the coupling $g_{0}$ saturates to  the value $g_{s}$, given by Eq.~(\ref{coupling-saturated}). Saturation is faster for more elongated particles possessing hot spots near their tips, in which case the QE-plasmon coupling reaches 90\% of its saturated value for $a_{1}/a=2$.

Finally, let us compare the coupling parameters for large QE ensembles coupled to plasmons and to  cavity modes in the case when QEs saturate their  respective mode volumes.  While in plasmonic cavities, the QE-plasmon coupling  is strongly enhanced relative to the QE coupling to a cavity mode, this is not so for \textit{open} plasmonic systems since, in the latter case, only a small fraction of QEs close to the metal surface participates in the energy exchange with the plasmon mode. Indeed, let us assume that QEs distributed uniformly, with concentration $n$, inside a microcavity. The QE-cavity coupling is given by Eq.~(\ref{coupling-cet}) with  the cavity mode volume ${\cal V}_{cav}$ given by \cite{khitrova-nphys06}
\begin{equation}
\label{mode-volume-cavity}
\frac{1}{{\cal V}_{\rm cav}}
= \frac{\varepsilon(\bm{r})\bm{E}^{2}(\bm{r})}{\int \! dV  \varepsilon (\bm{r})\bm{E}^{2}(\bm{r})},
\end{equation}
where $\varepsilon(\bm{r})$ is the cavity dielectric function. After  averaging Eq.~(\ref{mode-volume-cavity}) over the cavity volume $V_{\rm cav}$, we obtain ${\cal V}_{\rm cav}\simeq V_{\rm cav}$, and the saturated coupling   for microcavities takes the form
\begin{equation}
\label{coupling-saturated-cavity}
g_{s}^{\rm cav} 
=\sqrt{\frac{2\pi\mu^{2}n\omega_{m}}{3\hbar}}.
\end{equation}
Comparing $g_{c}^{\rm cav}$ to Eq.~(\ref{coupling-saturated2}), we conclude that, for similar QE concentrations, the saturated coupling for open plasmonic system is of the same order as  for microcavities, implying that large Rabi splittings observed in open plasmonic systems are likely due to  high  QE concentrations.

%%%%%%%%%%%%%%%%%%%%%%%%%%%%
%\section{Conclusions}

In conclusion, we developed a model for exciton-plasmon coupling  based on a microscopic picture of energy exchange that drives the transition to strong coupling regime. Plasmonic correlations between  QEs give rise to a collective state that transfers its energy cooperatively to a resonant plasmon mode at a rate  equal to the sum of individual QE-plasmon ET rates. This CET rate, along with the plasmon decay rate,  determines the QE-plasmon coupling and the energy-exchange frequency in the strong coupling regime. By defining accurately the plasmon mode volume, we have shown that, for a QE near a sharp metal tip, the QE-plasmon coupling is enhanced by the factor $\sqrt{\varepsilon'(\omega_{m})\lambda^{3}/V_{\rm met}}$ relative to the QE  coupling to a cavity mode, where $V_{\rm met}$ is the fraction of metal volume that largely confines the plasmon field.
For an ensemble of $N$ QEs placed in a region with weakly varying plasmon LDOS, the QE-plasmon coupling exhibits cavity-like scaling $g\propto \sqrt{N}$. However, in open plasmonic systems,  the ensemble QE-plasmon coupling saturates to a universal value $g_{s}$ that does not depend on the plasmonic system  size and shape, except indirectly via the plasmon frequency. Finally, we compared the coupling parameters for QEs interacting with plasmon and cavity modes and found that, for similar QE concentrations,   open plasmonic systems offer no significant enhancement, implying that the observed large Rabi splittings are likely due to high QE concentrations in such systems.

\acknowledgments

%%%%%%%%%%%%%%%%%%%%%%%%%%%%%%%
This work was supported  in part by the National Science Foundation under Grants  No. DMR-1610427,    No. DMR-1826886  and   No. HRD-1547754.

%%%%%%%%%%%%%%%%%%%%%%%%%%%%%%%%%%%%%

\end{document}